\documentclass{ws-ijmpb}

\begin{document}

\markboth{T. Tohyama and K. Tsutsui}
{Spectral weight of resonant inelastic x-ray scattering in doped cuprates}

%
\catchline{}{}{}{}{}
%

\title{SPECTRAL WEIGHT OF RESONANT INELASTIC X-RAY SCATTERING IN DOPED CUPRATES: EFFECT OF CORE-HOLE LIFETIME
}

\author{TAKAMI TOHYAMA}

\address{Department of Applied Physics, Tokyo University of Science, Tokyo 125-8585, Japan\\
tohyama@rs.tus.ac.jp}

\author{KENJI TSUTSUI
}

\address{Synchrotron Radiation Research Center, National Institutes for Quantum and Radiological Science and Technology, Hyogo 679-5148, Japan
\\
tutui@spring8.or.jp
}

\maketitle

\begin{history}
\today

\end{history}

\begin{abstract}
We examine the effect of core-hole lifetime on the spectral weight of resonant inelastic x-ray scattering (RIXS) in hole-doped cuprates. We calculate the spectral weight by using exact diagonalization technique for a $4\times 4$ doped Hubbard lattice and find that spin-flip channel detecting single-magnon excitation is less sensitive to the core-hole lifetime while in non-spin-flip channel the spectral weight is strongly dependent on the lifetime. In the latter, charge and two-magnon excitations predominately contribute to RIXS for short and long core-hole lifetimes, respectively. For a realistic value of the core-hole lifetime in cuprates, both the charge and two-magnon excitations are expected to contribute to non-spin-flip channel in RIXS when the incident-photon energy is tuned to the main peak of x-ray absorption spectrum.
\end{abstract}

\keywords{RIXS; cuprates; Hubbard model.}

\section{Introduction}

Resonant inelastic x-ray scattering (RIXS) experiments tuned to Cu $L$ edge have provided a lot of new insights about spin dynamics in spin-flip channel\cite{Ament2011,Dean2015} as well as charge dynamics in non-spin-flip channel\cite{Ishii2014} in cuprate superconductors. The two channels are selected predominately by the polarization dependence of incident photon. Not only such polarization dependence but also incident-photon energy dependence of the RIXS spectrum gives us useful information on the electronic states of hole-doped cuprates. In particular, non-spin-flip channel exhibits a fluorescence-like shift of spectral weight with increasing the incident-photon energy.\cite{Guarise2014,Minola2015,Huang2016}
Such a fluorescence-like shift has been interpreted as a consequence of a change of spectral weight distribution from two-magnon excitations appearing when the incident-phonon energy is tuned to the main peak of x-ray absorption spectrum (XAS) to charge excitations of hole carriers appearing when the incident-phonon energy is tuned for to a satellite structure in XAS.\cite{Tsutsui2016}

The spectral weight in Cu $L$-edge RIXS is also dependent on the lifetime of core hole in the intermediate state.\cite{Ament2011} Expanding the resolvent of the intermediate state in terms of the core-hole lifetime, i.e., ultrashort core-hole lifetime expansion,\cite{Brink2006,Ament2007} we may apply a perturbation theory for the RIXS spectrum. If the core-hole lifetime is short enough, spectral weight in spin-flip (non-spin flip) channel is represented by the dynamical spin (charge) structure factor, which is called fast collision approximation.\cite{Ament2009} However, spectral weights have been reported to be strongly sensitive to the order of the expansion.\cite{Jia2016} This shortcoming also leads to alternative way to construct an effective low-energy operator for RIXS.\cite{Haverkort2010} 

In this paper, we examine the effect of core-hole lifetime on RIXS spectral weight by using unbiased numerical exact diagonalization for a small two-dimensional Hubbard lattice representing doped cuprates. We find that spin-flip channel is relatively insensitive to the core-hole lifetime. In contrast, spectral weights for non-spin-flip channel depend on the value of the core-hole lifetime: charge and two-magnon excitations predominately contribute to RIXS for short and long core-hole lifetimes, respectively.

This paper is organized as follows. The Hubbard model and RIXS spectra decomposed into spin-flip and non-spin-flip channels are introduced in Sec.~\ref{Sec2}. In Sec.~\ref{Sec3}, we calculate the core-hole lifetime dependence of RIXS spectra for both spin-flip and non-spin-flip channels, and discuss their implications. A summary is given in Sec.~\ref{Sec4}.

\section{Hubbard model and RIXS spectrum}
\label{Sec2}

In order to describe $3d$ electrons in the CuO$_2$ plane, we take a single-band Hubbard model given by
\begin{equation}
H_\mathrm{3d}=-t\sum_{i\delta\sigma} c^\dagger_{i\sigma} c_{i+\delta\sigma}  + U\sum_i n_{i\uparrow}n_{i\downarrow},
\label{singleH}
\end{equation}
where $c^\dagger_{i\sigma}$ is the creation operator of an electron with spin $\sigma$ at site $i$; number operator $n_{i\sigma}=c^\dagger_{i\sigma}c_{i\sigma}$; $i+\delta$ represents the four first nearest-neighbor sites around site $i$; and $t$ and $U$ are the nearest-neighbor hopping and on-site Coulomb interaction, respectively. Next-nearest-neighbor hoppings are neglected for simplicity.

Tuning polarization of incident and outgoing photons in RIXS, we can separate an excitation with the change of total spin by one, i.e., spin-flip channel with the change of spin $\Delta S=1$ and an excitation with no change of total spin, i.e., non-spin-flip channel with $\Delta S=0$.\cite{Haverkort2010,Igarashi2012,Kourtis2012,Tohyama2015}
The two excitations can be defined as
\begin{eqnarray}\label{IDS0}
I^{\Delta S=0}_\mathrm{RIXS} \left(\mathbf{q}, \omega \right) &=& \sum\limits_f \left| \left\langle f \right|N^j_\mathbf{q} \left| 0 \right\rangle \right|^2 \delta \left( \omega  - E_f + E_0 \right),\\
\label{IDS1}
I^{\Delta S=1}_\mathrm{RIXS} \left(\mathbf{q}, \omega \right) &=& \sum\limits_f \left| \left\langle f \right|S^j_\mathbf{q} \left| 0 \right\rangle \right|^2 \delta \left( \omega  - E_f + E_0 \right),
\end{eqnarray}
with $S^j_\mathbf{q}=(B^j_{\mathbf{q}\uparrow\uparrow}-B^j_{\mathbf{q}\downarrow\downarrow})/2$, $N^j_\mathbf{q}=B^j_{\mathbf{q}\uparrow\uparrow}+B^j_{\mathbf{q}\downarrow\downarrow}$, and
\begin{equation}\label{Bqw}
B^j_{\mathbf{q}\sigma'\sigma}=\sum_l e^{-i\mathbf{q}\cdot\mathbf{R}_l} c_{l\sigma'}\frac{1}{\omega_\mathrm{i}-H_l^j+E_0+i\Gamma} c^\dagger_{l\sigma},
\end{equation}
where $\left|0 \right\rangle$ ($\left|f \right\rangle$) represents the ground state (the final state) with energy $E_0$ ($E_f$) in the Hubbard model (\ref{singleH}); $j$ is the total angular momentum of Cu$2p$ with either $j=1/2$ or $j=3/2$; $\Gamma$ is the inverse of core-hole lifetime; and $H_l^j=H_{3d}-U_\mathrm{c} \sum_\sigma n_{l\sigma} + \varepsilon_j$ with $U_\mathrm{c}$ and $\varepsilon_j$ being the Cu $2p$-$3d$ Coulomb interaction and energy level of Cu $2p$, respectively. Here, we assume the presence of a Cu$2p$ core hole at site $l$. $\mathbf{R}_l$ is the position vector at site $l$. $\omega_\mathrm{i}$ in the denominator of Eq.~(\ref{Bqw}) represents the incident-photon energy tuned to the main peak position of XAS calculated by 
\begin{equation}\label{XAS}
I^\mathrm{XAS}\left(\omega \right) = -\frac{1}{\pi}\mathrm{Im} \left\langle 0 \right| \sum_{l\sigma} c_{l\sigma}\frac{1}{\omega-H_l^j+E_0+i\Gamma} c^\dagger_{l\sigma}\left| 0 \right\rangle.
\end{equation}

When $\Gamma$ is much larger than the remaining terms in the denominator of Eq.~(\ref{Bqw}), $S^j_\mathbf{q}$ and $N^j_\mathbf{q}$ reduce to $S^z_\mathbf{q}=\sum_l e^{-i\mathbf{q}\cdot\mathbf{R}_l} S^z_l$ and $N_\mathbf{q}=\sum_l e^{-i\mathbf{q}\cdot\mathbf{R}_l} N_l$, respectively, with the $z$ component of the spin operator $S^z_l$ and electron-number operator $N_l$. This is nothing but the first-collision approximation, where Eqs.~(\ref{IDS0}) and (\ref{IDS1}) read the dynamical charge structure factor,
\begin{equation}
N(\mathbf{q},\omega)=\sum\limits_f \left| \left\langle f \right|N_\mathbf{q} \left| 0 \right\rangle \right|^2 \delta \left( \omega  - E_f + E_0 \right), \label{Nqw}
\end{equation}
and the dynamical spin structure factor,
\begin{equation}
S(\mathbf{q},\omega)=\sum\limits_f \left| \left\langle f \right|S^z_\mathbf{q} \left| 0 \right\rangle \right|^2 \delta \left( \omega  - E_f + E_0 \right), \label{Sqw}
\end{equation}
respectively.

In order to calculate Eqs.~(\ref{IDS0}), (\ref{IDS1}),  (\ref{XAS}),(\ref{Nqw}), and (\ref{Sqw}), we use a Lanczos-type exact diagonalization technique on a $4\times 4$ cluster under periodic boundary conditions.
We consider hole doping with hole concentration $x=1/8$, i.e., 2-hole doped $4\times 4$ cluster. We take $U/t=U_\mathrm{c}/t=10$ and $\Gamma=t$ for XAS but $\Gamma$ in RIXS is regarded as a parameter to examine the effect of the core-hole lifetime on RIXS spectra.

\section{Core-hole lifetime dependence of RIXS spectra}
\label{Sec3}

\begin{figure}[t]
\centerline{\psfig{file=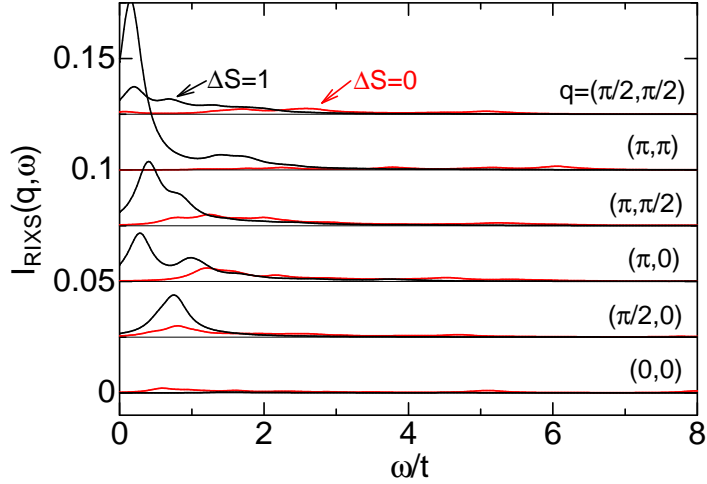,width=3.65in}}\vspace*{8pt}
\caption{RIXS spectra at various $\mathbf{q}$ on the $4\times 4$ single-band Hubbard cluster with $U=U_\mathrm{c}=10t$ and $\Gamma=t$ for hole doping with hole concentration $x=2/18=0.125$. Black and red colors represent $\Delta S=1$ and $\Delta S=0$ excitations, respectively. The solid curves are obtained by performing a Lorentzian broadening with a width of $0.2t$. The incident photon energy is set to the main peak of absorption spectrum.}
\label{fig1}
\end{figure}

Before examining $\Gamma$ dependence, we calculate RIXS spectra for all of $\mathbf{q}$ defined in the $4\times 4$ cluster. Figure~\ref{fig1} shows both cases of $\Delta S=1$ and $\Delta S=0$ for $\Gamma=1$. The spectra for $\Delta S=1$ originating from single-magnon excitations are distributed lower in energy than those for $\Delta S=0$ as expected. In particular, at $\mathbf{q}=(\pi,\pi)$ the spectra are well separated between low-energy region for $\Delta S=1$ and high-energy region up to $\omega=6t$ for $\Delta S=0$. At $\mathbf{q}=(0,0)$, there is no spectral weight for $\Delta S=1$, while there appears small weight around $\omega=0.7t$ for $\Delta S=0$. This contrasting behavior indicates that spin-flip channel resembles to $S(\mathbf{q},\omega)$ whose weight at $\mathbf{q}=(0,0)$ is zero, while non-spin-flip channel does not necessarily represent $N(\mathbf{q},\omega)$ having no weight at $\mathbf{q}=(0,0)$.

\begin{figure}[t]
\centerline{\psfig{file=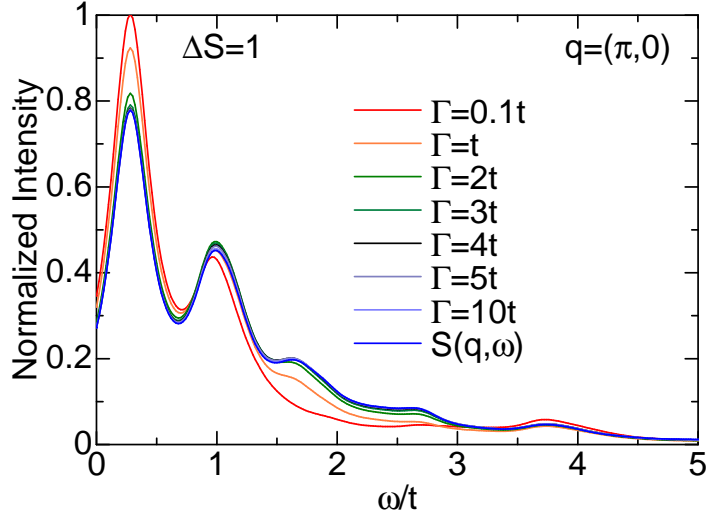,width=3.65in}}
\vspace*{8pt}
\caption{The inverse core-hole lifetime $\Gamma$ dependence of spin-flip $\Delta S=1$ RIXS spectra at $\mathbf{q}=(\pi,0)$ on the $4\times 4$ single-band Hubbard cluster with $U=U_\mathrm{c}=10t$ for hole doping with hole concentration $x=2/18=0.125$. Each color corresponds to a value of $\Gamma$ denoted in the panel except for the blue solid line representing the dynamical spin structure factor $S(\mathbf{q},\omega)$.}
\label{fig2}
\end{figure}

In order to examine $\Gamma$ dependence, we focus on RIXS spectra at $\mathbf{Q}=(\pi,0)$. Figure~\ref{fig2} shows the $\Gamma$ dependence for $\Delta S=1$. The spectrum at large $\Gamma$ is the same as $S(\mathbf{q},\omega)$ as expected. We find that basic spectral features of $S(\mathbf{q},\omega)$ remain even for small $\Gamma$ down to $\Gamma=0.1t$. This means that spin-flip channel in RIXS for cuprates predominately describes single-magnon excitations expressed by the dynamical spin structure factor.

\begin{figure}[t]
\centerline{\psfig{file=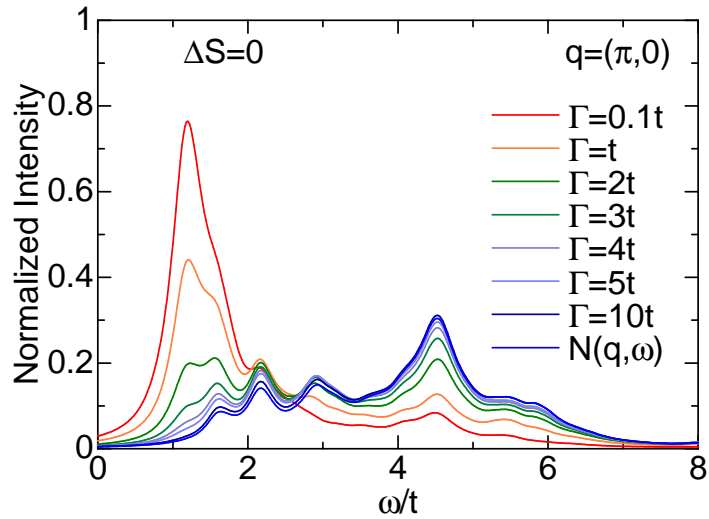,width=3.65in}}
\vspace*{8pt}
\caption{The inverse core-hole lifetime $\Gamma$ dependence of  non-spin-flip $\Delta S=0$ RIXS spectra at $\mathbf{q}=(\pi,0)$ on the $4\times 4$ single-band Hubbard cluster with $U=U_\mathrm{c}=10t$ for hole doping with hole concentration $x=2/18=0.125$. Each color corresponds to a value of $\Gamma$ denoted in the panel except for the blue solid line representing the dynamical spin structure factor $N(\mathbf{q},\omega)$.}
\label{fig3}
\end{figure}

In contrast to the spin-flip $\Delta S=1$ case, non-spin-flip channel with $\Delta S=0$ exhibits strong $\Gamma$ dependence. Figure~\ref{fig3} shows the $\Gamma$ dependence for $\Delta S=0$. The spectrum at large $\Gamma$ is again the same as $N(\mathbf{q},\omega)$ as expected. However, with decreasing $\Gamma$, spectral distribution deviates from $N(\mathbf{q},\omega)$ with the reduction of spectral weight, and at the same time low-energy weight at $\omega=1.2t$ largely increases. At the smallest value of $\Gamma$, the low-energy peak dominates spectrum. When $\Gamma$ is comparable with the remaining terms in the denominator of Eq.~(\ref{Bqw}), intersite operators emerge as effective operators in Eq.~(\ref{Bqw}), in addition to the on-site charge operator $N_i$.\cite{Jia2016} In non-spin-flip channel, a two-magnon-type operator is easily expected to contribute to $N^j_\mathbf{q}$ in Eq.~(\ref{IDS0}). In fact, comparing with dynamical two-magnon correlation function at $\mathbf{q}=(\pi,0)$ (not shown), we find that the low-energy peak corresponds to a two-magnon excitation.

For a realistic parameter of $\Gamma (\sim t)$ for cuprates and $\omega_\mathrm{i}$ tuned to the main peak of XAS, a large amount of two-magnon contribution exists in addition to charge excitations.\cite{Tsutsui2016} Therefore, it is important to carefully distinguish both contributions. In fact, the presence of these two components contribute to a florescence-like $\omega_\mathrm{i}$ dependence of non-spin-flip channel in RIXS for cuprates.\cite{Tsutsui2016}

\section{Summary}
\label{Sec4}
We have investigated the effect of core-hole lifetime on RIXS spectral weight by using unbiased numerical exact diagonalization for the $4\times 4$ two-dimensional doped Hubbard lattice. We find that spin-flip channel is less insensitive to the core-hole lifetime. In contrast, non-spin-flip channel, where both two-magnon excitation and charge excitation emerge, depends on the value of the core-hole lifetime: for short lifetime charge excitations are dominated, while for long lifetime two-magnon excitations control low-energy excitations. It is reasonable to detect two-magnon excitations for long core-hole lifetime, since an electron excited to the Hubbard band from core level can propagate the system during the core-hole lifetime, leading to effective intersite excitations like two magnons. We should be careful for the fact that  both charge and two-magnon excitations are expected to contribute to RIXS spectra for a realistic value of the core-hole lifetime for cuprates.

\section*{Acknowledgements}
This work was supported by the Japan Society for the Promotion of Science, KAKENHI (Grants No. 26287079, No. 15H03553, and No. 16H04004) and Creation of new functional devices and high-performance materials to support next-generation industries (GCDMSI) to be tackled by using post-K computer and by MEXT HPCI Strategic Programs for Innovative Research (SPIRE) (hp160222, hp170274) from Ministry of Education, Culture, Sports, Science, and Technology.

\end{document}